\documentclass{article}
\usepackage{color}
\usepackage{amsmath}
\usepackage{amsthm}
\usepackage{titlesec}
\usepackage{indentfirst}
\usepackage{cite}
\begin{document}

\title{{\bf Sharing a quantum secret without a trusted party}}
\author{\bf {Qin Li$^{1,2,}$\footnote{Corresponding author: Qin Li \vskip 1pt  \hspace {4pt} { E-mail address:}
liqin805@163.com (Q. Li)} , Dong Yang Long$^1$, W. H. Chan$^2$, Dao Wen Qiu$^1$}\\
\footnotesize{$^1$\sl Department of Computer Science, Sun Yat-sen
University, Guangzhou 510006 China}\\
\footnotesize{$^2$\sl Department of Mathematics, Hong Kong Baptist
University, Kowloon, Hong Kong, China} \\
\footnotesize{}}
\date{ }
\maketitle

{\flushleft{\bf Abstract}}\hskip 10pt In a conventional quantum
$(k,n)$ threshold scheme, a trusted party shares a secret quantum
state with $n$ participants such that any $k$ of those participants
can cooperate to recover the original secret, while fewer than $k$
participants obtain no information about the secret. In this paper
we show how to construct a quantum $(k,n)$ threshold scheme without
the assistance of a trusted party, who generates and distributes
shares among the participants. Instead, each participant chooses his
private state and contributes the same to the determination of the
final secret quantum state.

\leftline{{\bf Keywords} \hskip 10pt Quantum secret sharing $\cdot$
Quantum cryptography $\cdot$ Quantum information processing $\cdot$
} \leftline{Quantum communication}

 \leftline{{\bf PACS} \hskip 10pt 03.67.Dd}

\noindent\textbf{1 Introduction}

Suppose that $n$ shareholders who are not mutually trusted want to
share a password, with which they can open the vault and get access
to some confidential documents of the company. It should be done in
such a way that any $k$ of those shareholders have the ability to
reconstruct the password, while fewer than $k$ shareholders cannot.
The solutions vary under two different situations. If there exists a
trusted party who generates and distributes suitable shares among
the shareholders, then the problem could be addressed by classical
$(k,n)$ threshold schemes independently introduced by Blakley
\cite{bla} and Shamir \cite{sha}. And if there does not exist such a
party trusted by all of the shareholders, then classical $(k,n)$
threshold schemes without the assistance of any trusted party came
to rescue \cite{is,ped,jmo}.

With the emergence of quantum computation and quantum communication,
it is natural to consider the quantum counterparts of secret sharing
schemes. Hillery \emph{et al}. showed how to implement a classical
threshold scheme using Greenberger-Horne-Zeilinger (GHZ) states in
the presence of eavesdroppers, and also showed how to share an
unknown qubit between two participants such that only the
collaboration of two participants could reconstruct the original
qubit \cite{hbb}. Karlsson \emph{et al}. presented a secret sharing
scheme based on two-particle quantum entanglement, in which quantum
information is sent from a sender Trent to Alice and Bob so that
both persons are needed to obtain the information, and showed that
it could be extended to a special quantum $(k,n)$ threshold scheme
based on multi-particle entangled states \cite{kki}. Cleve \emph{et
al}. gave an efficient construction of more general quantum $(k,n)$
threshold schemes, where a trusted party can share an unknown
quantum state with $n$ participants such that $k$ participants are
necessary and sufficient to reconstruct the original secret quantum
state \cite{cgl}. Thereafter quantum secret sharing has been an
active research field and many quantum secret sharing schemes using
various techniques have been proposed
\cite{dg,sb,nmi,gg,lsb,xldp,toi,zjz,dzl,ti,zjz2,ylh}.

Whereas most previous quantum $(k,n)$ threshold schemes have been
considered with the assistance of a trusted party
\cite{hbb,kki,cgl,dg,sb,nmi,gg,lsb,xldp,toi,zjz,dzl,ti,zjz2,ylh},
here we consider the problem of sharing a quantum secret without
having a trusted party. Actually, for lots of applications such as
conference key agreement and distributed computing with faulty
processors, quantum secret sharing without a trusted party is a
powerful tool. Also, as referred by Ingemarsson and Simmons
\cite{is}, the situation that there is nobody trusted by all of the
participants is more common in commercial and/or international
applications. In this case, previously proposed quantum secret
schemes do not play the role well. Therefore, in this paper we give
a basic construction for sharing a quantum secret in the absence of
a trusted party and illustrate its feasibility mainly by improving
the quantum $(k, n)$ threshold scheme presented by Cleve \emph{et
al}. \cite{cgl} to eliminate the need of the trusted party. The
novelty is that a trusted party is unnecessary and each participant
acts for his own benefit.

\noindent\textbf{2 Quantum $(k,n)$ threshold schemes without a
trusted party}

In this section, we show how to solve the problem of sharing a
quantum secret without the aid of any trusted party, drawing ideas
from classical counterparts \cite{is,ped,jmo}. The problem could be
stated more clearly as follows:
\begin{itemize}
\item all of the participants choose their own private quantum states, which
must be made available in sequence for the final secret quantum
state to be generated;

\item all of the participants contribute identically to determining
the quantum secret;

\item the participants share the secret quantum state in a way that the collaboration
of any $k$ out of $n$ participants can recover the secret quantum
state, but the collaboration of any $k-1$ or fewer participants
obtain no information about it.
\end{itemize}
Such a scheme is called a \emph{quantum $(k,n)$ threshold scheme
without a trusted party}. In such a scheme, each participant acts
for his own benefit and need not trust a single party unless at
least $k$ participants work together. In addition, notice $n<2k$ for
a quantum $(k,n)$ threshold scheme without a trusted party, which is
guaranteed by quantum no-cloning theorem \cite{woot}.

The basic method used for constructing a quantum $(k,n)$ threshold
scheme without a trusted party is introduced in the following.

(1) Each participant $P_i$ $(i=0,1,\cdots,n-1)$ randomly selects his
private quantum state $\rho_i$ which can be a component of the
quantum secret, or belong to the same domain as that of the secret
quantum state to be reconstructed. The specific form of the quantum
secret can vary with different applications and should be defined
before implementing the protocol.

(2) $P_i$ acts as a trusted party and splits his private quantum
state $\rho_i$ into $n$ shares $\rho_{ij}$ $(j=0,1,\cdots,n-1)$
using a conventional quantum $(k,n)$ threshold scheme such as
\cite{cgl}, and then sends each share $\rho_{ij}$ to the participant
$P_j$ in sequence (particularly $P_i$ keeps $\rho_{ii}$).

(3) When $P_i$ has received the share $\rho_{ji}$ from the
participants $P_j$, he can do nothing or implement corresponding
operations on $\rho_{ji}$ and his quantum registers. Note that such
operations are agreed before the protocol and vary with the
conventional quantum $(k,n)$ threshold schemes used. If $P_i$ has
received the shares $\rho_{ji}$ ($j=0,1,\cdots,n-1$) and implemented
the corresponding operations by some time (the time limit is agreed
before the protocol), he announces that he has completed his
actions; else if $P_i$ has not get the share from some party $P_l$,
he also reveals this fact.

(4) If at least $k$ parties have not received the shares from some
party $P_l$ before the time limit, they consider $P_l$ a quitter and
not a real player. If the number of quitters is not exceeding $n-k$,
the protocol cannot be affected and just the value of $n$ decreases;
otherwise the protocol aborts. After all of the real participants
have announced, any $k$ participants can cooperate to obtain the
final secret quantum state $\rho_s$ which each participant's private
state $\rho_i$ contributes equally to by a preconcerted way, while
less than $k$ participants obtain no information about $\rho_s$
except their own private states.

Note that we should assume that each participant must logically be
willing to share the private quantum state chosen by himself with
other participants and contribute the same to determining the final
secret quantum state. Otherwise a malicious participant who sends an
improper quantum states can make $k$ participants reconstruct a
wrong quantum secret and thus make the protocol abort, even though
doing such things is of no use for him. In addition, the security of
such a quantum $(k,n)$ threshold scheme without having a trusted
party largely depends on the conventional quantum $(k,n)$ threshold
scheme used by each participant to protect his private state among
the other participants before the recovery phase.

\noindent\textbf{3 Concrete quantum $(k,n)$ threshold schemes
without a trusted party}

In order to clarify the feasibility of the construction of a quantum
$(k,n)$ threshold scheme without a trusted party introduced by us,
we give a couple of schemes. Besides, it is only necessary for us to
consider real players' actions, since the construction method has
ensured that if at most $n-k$ players drop out midway, the protocol
will not be affected.

one scheme can be achieved as follows just by utilizing the
operations included in a conventional quantum $(k,n)$ threshold
scheme such as that in \cite{hbb,kki,cgl}: each participant shares
the private state chosen by himself with the other participants
using a conventional quantum $(k, n)$ threshold scheme; if a
participant has obtained the shares and performed accordingly, he
makes an announcement that his task has been finished; then any $k$
players recover each participant's private state using the
conventional recovery procedure; and such $k$ players obtain the
final secret from all of the recovered private states by
implementing the agreed operations. Obviously, the processes of this
scheme can accord with that of the proposed construction method and
the trusted party is removed indeed. Nevertheless, the conventional
recovery procedure has to be implemented for $n$ times and then the
negotiated operations should be applied on $n$ private states to
obtain a secret.

Another concrete scheme is based on the quantum $(k,n)$ threshold
scheme presented by Cleve \emph{et al}. \cite{cgl}.  We improve it
to remove the use of the trusted party. The improved scheme can
allow all of the participants to choose their own private state and
have the same influence on determining the final quantum secret.

Note $n<2k$ and that the dimension of each share can be bounded
above by $2\max(2k-1,m)$ through efficient quantum operations, where
$m$ is the dimension of the private quantum state to be encoded, the
same as that in \cite{cgl}. In addition, we just need to consider
the special case where $n=2k-1$, since a quantum $(k,n)$ threshold
scheme with $n<2k-1$ can be obtained by discarding $2k-1-n$ shares
from any $(k,2k-1)$ threshold scheme with $n>k$ \cite{cgl}. Our
improved scheme without the aid of the trusted party includes three
parts: scheme setup, secret quantum state generation, and secret
quantum state reconstruction.

 \vspace{5pt}\noindent{3.1 Scheme setup} \vspace{5pt}

Given $k$, $n$ and $m$, find a suitable prime $q$ satisfying
$\max(n,m)\leq q \leq 2\max(n,m)$ (which is always possible
according to Bertrand's postulate \cite{az}) and set a finite field
$\textbf{F}=\textbf{Z}_q$. For $i=0,1,\cdots,n-1$, let
$c_i=(c_{i,0},c_{i,1},\cdots,c_{i,{k-1}})\in \textbf{F}^k$, define
the polynomial as
$p_{c_i}(t)=c_{i,0}+c_{i,1}t+\cdots+c_{i,{k-1}}t^{k-1}$, and let
$x_i\in \textbf{F}$ and each $x_i$ should be different from each
other. Then a $q$-ary quantum state which is defined on basis states
$|s_i\rangle$ ($s_i\in \textbf{F}$) could be encoded by the linear
mapping as
\begin{eqnarray}
|s_i\rangle\rightarrow \sum_{
\begin{array}{c}
c_i\in \textbf{F}^k, c_{i,{k-1}}=s_i
\end{array}}
|p_{c_i}(x_0),p_{c_i}(x_1),\cdots,p_{c_i}(x_{n-1})\rangle
\label{eq:one}.
\end{eqnarray}

Define the resulting state of adding a $q$-ary quantum basis state
$|s_i\rangle$ to the quantum register in a $q$-ary quantum basis
state $|s_j\rangle$ as $|s_i+s_j\rangle|s_i\rangle$ ($s_i,s_j \in
\textbf{F}$), which can be encoded by the linear mapping as

\begin{eqnarray}
\sum_{
\begin{array}{c}
c_i,c_j\in \textbf{F}^k\\
c_{i,k-1}=s_i,c_{j,k-1}=s_j
\end{array}}
|p_{c_i}(x_0)+p_{c_j}(x_0),\cdots,p_{c_i}(x_{n-1})+p_{c_j}(x_{n-1})\rangle|p_{c_i}(x_0),\cdots,p_{c_i}(x_{n-1})\rangle
\label{eq:two}.
\end{eqnarray}

Also define the operation applying an invertible $l\times l$ matrix
$M$ to a sequence of $l$ quantum registers as applying the mapping
\begin{equation}
|(y_0,y_1,\cdots,y_{l-1})\rangle\rightarrow
|(y_0,y_1,\cdots,y_{l-1})M\rangle. \label{eq:three}
\end{equation}
For $z_0,z_1,\cdots,z_{l-1}\in \textbf{F}$, introduce the $l\times
l$ Vandermonde matrix $[V_l(z_0,z_1,\cdots,z_{l-1})]_{ij}=z_j^i$
$(i,j\in \{0,1,\cdots,l-1\})$. And notice that
\begin{eqnarray}
& &
|(c_{i,0},c_{i,1},\cdots,c_{i,l-1})V_l(z_0,z_1,\cdots,z_{l-1})\rangle
\nonumber \\
&=& |p_{c_i}(z_0),p_{c_i}(z_1),\cdots,p_{c_i}(z_{l-1})\rangle,
\label{eq:four}
\end{eqnarray}
where $c_i=(c_{i,0},c_{i,1},\cdots,c_{i,l-1})\in \textbf{F}^l$.

In addition, suppose that each participant $P_i$
$(i=0,1,\cdots,n-1)$ owns a local quantum register $R_i$ and $n$
common quantum registers $C_{i,j}$ $(j=0,1,\cdots,n-1)$ and every
register is in an initial state $|0\rangle$ ($0\in \textbf{F}$).

\vspace{5pt}\noindent{3.2 Secret quantum state generation}
\vspace{5pt}

In this phase, $P_i$ ($i=0,1,\cdots,n-1$) will adopt a method of
encoding by employing a polynomial of degree $k-1$ denoted as
$p_{c_i}(t)=c_{i,0}+c_{i,1}t+\cdots+c_{i,{k-1}}t^{k-1}$, where
$c_i=(c_{i,0},c_{i,1},\cdots,c_{i,{k-1}})\in \textbf{F}^k$.

$P_i$ randomly chooses and prepares his own private $q$-ary quantum
state $|s_i\rangle$ ($s_i\in \textbf{F}$) and uses the encoding
denoted as (\ref{eq:one}) to obtain
\begin{eqnarray}
\sum_{
\begin{array}{c}
c_i\in \textbf{F}^k, c_{i,{k-1}}=s_i
\end{array}}
|p_{c_i}(x_0),p_{c_i}(x_1),\cdots,p_{c_i}(x_{n-1})\rangle.
\label{eq:five}
\end{eqnarray}
Note that the above operation implies that the quantum state
$|s_i\rangle$ is split into $n$ shares which are stored in the
quantum registers $C_{i,j}$ $(j=0,1,\cdots,n-1)$. Then $P_i$ sends
the share in the register $C_{i,j}$ to $P_j$ for $j=0,1,\cdots,n-1$
(including himself) sequently in an authenticated way.

When $P_i$ receives a share, he adds the received state to the
quantum register $R_i$. And if $P_i$ has obtained the shares from
all of the participants and implemented all the corresponding
operations, he announces that his actions have been finished. After
all of the participants have announced, it is not difficult to
obtain that $n+n^2$ quantum registers
$R_0,R_1,\cdots,R_{n-1},C_{0,0},C_{0,1},\cdots,C_{{n-1},{n-1}}$ are
in a global state
\begin{eqnarray}
\sum_{
\begin{array}{c}
c_i\in \textbf{F}^k,c_{i,k-1}=s_i\\
\text{for } i=0,\cdots,n-1
\end{array}}
|\sum\limits_{i = 0}^{n - 1}p_{c_i}(x_0),\cdots,\sum\limits_{i =
0}^{n -
1}p_{c_i}(x_{n-1})\rangle\otimes^{n-1}_{i=0}|p_{c_i}(x_0),\cdots,p_{c_i}(x_{n-1})\rangle.
\label{eq:six}
\end{eqnarray}

Then, let the final agreed secret quantum state be $|s\rangle$ (for
$s=\sum\limits_i s_i=\sum\limits_ic_{i,{k-1}}$ and $s\in\textbf{
F}$) if all the participants are required to provide $q$-ary quantum
basis states before the scheme. In this case, we shall see that the
final secret can be recovered using the conventional recovery
procedure only once. However, if players can be allowed to provide a
superposition of basis states, their private states may entangle
with the resulting state after performing operations on them. So the
final secret state should be defined in a different way, such as a
superposition of states
$|\sum\limits_{i=0}^{n-1}s_i\rangle\otimes^{n-1}_{i=0}|s_i\rangle
(s_i\in\textbf{F})$ or measuring the superposition state using the
computational basis in $F^{n+1}$ and taking the collapsed state
$|\sum\limits_{i=0}^{n-1}s_i\rangle$ in the first register as a
final secret. The form of the final secret can vary with the
specific applications and should be negotiated before implementing
the scheme.

\vspace{5pt}\noindent{3.3 Secret quantum state reconstruction}
\vspace{5pt}

In this part, we show that the collaboration of any $k$ participants
can reconstruct the agreed secret quantum state $|s\rangle$ by the
following steps. Suppose that the first $k$ participants, namely
$P_0,P_1,\cdots,P_{k-1}$, gather together and want to recover the
secret, so we obtain the information about the fist $k$ quantum
registers (that is, $R_0,\cdots,R_{k-1}$) and $kn$ common quantum
registers (namely, $C_{i,j}$ for $i=0,\cdots,n-1$ and
$j=0,\cdots,k-1$).

(1) Apply $V_k(x_0,x_1,\cdots,x_{k-1})^{-1}$ which represents the
inverse of $V_k(x_0,x_1,\cdots,x_{k-1})$ to the first $k$ quantum
registers $R_0,R_1,\cdots,R_{k-1}$. Then the global state of the
$n+n^2$ quantum registers is {\small
\begin{eqnarray}
\sum_{
\begin{array}{c}
c_i\in \textbf{F}^k,c_{i,k-1}=s_i\\
\text{for } i=0,\cdots,n-1
\end{array}}
|\sum\limits_{i = 0}^{n - 1}c_{i,0},\cdots,\sum\limits_{i = 0}^{n -
1}c_{i,{k-1}}\rangle|\sum\limits_{i = 0}^{n -
1}p_{c_i}(x_{k}),\cdots,\sum\limits_{i = 0}^{n -
1}p_{c_i}(x_{n-1})\rangle\otimes^{n-1}_{i=0}|p_{c_i}(x_0),\cdots,p_{c_i}(x_{n-1})\rangle.
\label{eq:seven}
\end{eqnarray}}

(2) Shift the first $k$ quantum registers by one to the right in
sequence by setting $R_0,R_1,\cdots,R_{k-1}$ to
$R_{k-1},R_0,\cdots,R_{k-2}$. At this time, the global state of the
$n+n^2$ quantum registers is {\scriptsize
\begin{eqnarray}
&&\sum_{
\begin{array}{c}
c_i\in \textbf{F}^k,c_{i,k-1}=s_i\\
\text{for }i=0,\cdots,n-1
\end{array}}
|\sum\limits_{i = 0}^{n - 1}c_{i,k-1},\sum\limits_{i = 0}^{n - 1}c_{i,0},\cdots,\sum\limits_{i = 0}^{n -
1}c_{i,{k-2}}\rangle|\sum\limits_{i = 0}^{n -
1}p_{c_i}(x_{k}),\cdots,\sum\limits_{i = 0}^{n -
1}p_{c_i}(x_{n-1})\rangle\otimes^{n-1}_{i=0}|p_{c_i}(x_0),\cdots,p_{c_i}(x_{n-1})\rangle\nonumber\\
&=& \sum_{
\begin{array}{c}
c_i\in \textbf{F}^k,c_{i,k-1}=s_i\\
\text{for }i=0,\cdots,n-1
\end{array}}
|s\rangle|\sum\limits_{i = 0}^{n - 1}c_{i,0},\cdots,\sum\limits_{i =
0}^{n - 1}c_{i,{k-2}}\rangle|\sum\limits_{i = 0}^{n -
1}p_{c_i}(x_{k}),\cdots,\sum\limits_{i = 0}^{n -
1}p_{c_i}(x_{n-1})\rangle\otimes^{n-1}_{i=0}|p_{c_i}(x_0),\cdots,p_{c_i}(x_{n-1})\rangle.
\label{eq:eight}
\end{eqnarray}}If the state in $R_0$ is a basis state $|s\rangle$ (for some $s\in
\textbf{F}$), it can be the final secret quantum state and the
recovery procedure has been done; otherwise we should continue the
recovery procedure. Actually, a participant usually offers a
superposition of basis states, so the register $R_0$ is entangled
with the other registers, since in (\ref{eq:eight}) the value of
$|s\rangle$ can be determined by any of the kets
\begin{eqnarray}
|\sum\limits_{i = 0}^{n - 1}c_{i,0},\cdots,\sum\limits_{i = 0}^{n -
1}c_{i,{k-2}}\rangle |\sum\limits_{i = 0}^{n -
1}p_{c_i}(x_{k}),\cdots,\sum\limits_{i = 0}^{n -
1}p_{c_i}(x_{n-1})\rangle\otimes^{n-1}_{i=0}|p_{c_i}(x_0),\cdots,p_{c_i}(x_{n-1})\rangle.\label{eq:nine}
\end{eqnarray}

(3) Apply $V_{k-1}(x_k,x_{k+1},\cdots,x_{n-1})$ to the quantum
registers $R_1,R_2,\cdots,R_{k-1}$ and add
$R_0\cdot(x_{k+i-1})^{k-1}$ to $R_i$ for all
$i\in\{1,2,\cdots,k-1\}$. And the global state of the $n+n^2$
quantum registers is {\scriptsize\begin{eqnarray} &&\sum_{
\begin{array}{c}
c_i\in \textbf{F}^k,c_{i,k-1}=s_i\\
\text{for }i=0,\cdots,n-1
\end{array}}
|s\rangle|\sum\limits_{i = 0}^{n -
1}p_{c_i}(x_{k}),\cdots,\sum\limits_{i = 0}^{n -
1}p_{c_i}(x_{n-1})\rangle|\sum\limits_{i = 0}^{n -
1}p_{c_i}(x_{k}),\cdots,\sum\limits_{i = 0}^{n -
1}p_{c_i}(x_{n-1})\rangle\otimes^{n-1}_{i=0}|p_{c_i}(x_0),\cdots,p_{c_i}(x_{n-1})\rangle.
\label{eq:ten}
\end{eqnarray}}

(4) The first $k$ parties perform the similar operations that were
applied to the registers $R_0,R_2,\cdots,R_{k-1}$ in steps (1), (2)
and (3) of the state reconstruction phase on the registers
$C_{i,0},C_{i,1},\cdots,C_{i,k-1}$ for $i=0,1,\cdots,n-1$. Then the
global state of the $n+n^2$ quantum registers should be
\begin{eqnarray}\sum_{
\begin{array}{c}
c_i\in \textbf{F}^k,c_{i,k-1}=s_i\\
\text{for }i=0,\cdots,n-1
\end{array}}
|s\rangle|\sum\limits_{i = 0}^{n -
1}p_{c_i}(x_{k}),\cdots,\sum\limits_{i = 0}^{n -
1}p_{c_i}(x_{n-1})\rangle|\sum\limits_{i = 0}^{n -
1}p_{c_i}(x_{k}),\cdots,\sum\limits_{i = 0}^{n -
1}p_{c_i}(x_{n-1})\rangle \nonumber\\ \otimes^{n-1}_{i=0}
(|s_i\rangle
|p_{c_i}(x_k),\cdots,p_{c_i}(x_{n-1})\rangle|p_{c_i}(x_k),\cdots,p_{c_i}(x_{n-1})\rangle).
\label{eq:eleven}
\end{eqnarray}

(5) Shift the registers possessed by the first participant $P_0$
(namely, $R_0,C_{0,0},\cdots,C_{n-1,0}$) to be first $n+1$
registers. For $i=(0,1,\cdots,k-1)$, since there is a unique array
$c_i\in\textbf{F}^k$ with $c_{i,{k-1}}=s_i$ such that
$p_{c_i}(x_{_{k+j-1}})=y_{i,j}$ for any $j\in\{1,2,\cdots,k-1\}$,
the state denoted as Eq. (\ref{eq:eleven}) can be rewritten as
\begin{eqnarray} |s\rangle\otimes^{n-1}_{i=0} |s_i\rangle\sum_{
\begin{array}{c}
c_i\in \textbf{F}^k,c_{i,k-1}=s_i\\
\text{for }i=0,\cdots,n-1
\end{array}}
|\sum\limits_{i = 0}^{n - 1}y_{i,1},\cdots,\sum\limits_{i = 0}^{n -
1}y_{i,k-1}\rangle|\sum\limits_{i = 0}^{n -
1}y_{i,1},\cdots,\sum\limits_{i = 0}^{n - 1}y_{i,k-1}\rangle
\nonumber\\
\otimes^{n-1}_{i=0}(|y_{i,1},\cdots,y_{i,k-1}\rangle|y_{i,1},\cdots,y_{i,k-1}\rangle).
\label{eq:twelve}
\end{eqnarray}
From the Eq. (\ref{eq:twelve}), if all the participants provide
$q$-ary quantum basis states, the state in $R_0$ is $|s\rangle$ (for
$s=\sum\limits_i s_i=\sum\limits_ic_{i,{k-1}}$ and $s\in\textbf{
F}$) and it can be the final agreed secret. If some players offer
superposition states of basis states, the state of the first $n+1$
registers (that is, $R_0,C_{0,0},\cdots,C_{n-1,0}$) is a
superposition of the states
$|\sum\limits_{i=0}^{n-1}s_i\rangle\otimes^{n-1}_{i=0}|s_i\rangle
(s_i\in\textbf{F})$ which can be defined as the final secret. We
also can measure the superposition state in the computational basis
in $F^{n+1}$ and the collapsed state
$|\sum\limits_{i=0}^{n-1}s_i\rangle$ in $R_0$ can be considered as a
final secret. Note that the the sum of the measurement outcomes of
$C_{0,0},\cdots,C_{n-1,0}$ should equate the value of $R_0$. If not,
there must be something wrong and the scheme should be carried out
again. The specific definition of the final secret can be adapted to
different applications and should be established before performing
the scheme.

In summary, the above two schemes where the trusted party is removed
demonstrate the feasibility of the construction method offered by
us. But in the first scheme, the conventional recovery procedure
needs to be implemented for $n$ times and then the agreed operations
on $n$ private states are needed to gain a secret . In the second
scheme, if all the participants are required to provide basis
states, the final secret can be reconstructed using the conventional
recovery procedure only once. Nevertheless, if participants are
allowed to offer superposition states, the conventional recovery
procedure should be employed for $n+1$ times to generate a secret.
In addition, the security of these two scheme largely relies on the
conventional quantum $(k,n)$ threshold scheme since each participant
needs it to protect his private state among the other participants
before the secret reconstruction phase.

 \vspace{5pt}\noindent\textbf{4 Conclusion} \vspace{5pt}

In this paper, we have given a basic construction of quantum $(k,n)$
threshold schemes without a trusted party and illustrated its
feasibility by offering two schemes. In contrast to previous
presented quantum secret sharing schemes
\cite{hbb,kki,cgl,dg,sb,nmi,gg,lsb,xldp,toi,zjz,dzl,ti,zjz2,ylh}, in
which a trusted party is always needed to assist in generating and
distributing shares among a group of participants, the schemes
suggested in this paper require no trusted party and thus might
widen the applicability of quantum threshold schemes to the
situation where there is no single party trusted by all of the
participants.

However, the construction method proposed needs the participants to
be logically honest during the secret generation phase since they
should provide a proper quantum state; otherwise a single
participant can disrupt the whole scheme and not be detected until
the secret reconstruction phase, even though doing so is of no use
for him. Although it is not quite reasonable to make such an
expectation, the main intention of this paper is to demonstrate the
necessity and feasibility of sharing a quantum secret without a
trusted party. Particularly, Michael Ben-Or recently suggested us
that the techniques used in quantum multiparty computation
\cite{cda,bcghs} may be exploited to propose quantum secret sharing
schemes in the absence of a trusted party where less than half of
the participants are not required to play honestly during the secret
generation phase. This invaluable advice shall build a bridge
between quantum secret sharing without the aid of a trusted party
and quantum multiparty computing. How to utilize them
comprehensively will be the further work.

\noindent\textbf{Acknowledgement}\hskip 10pt We would like to
appreciate Guang Ping He and Michael Ben-Or for useful suggestions.
We are also very grateful to the anonymous referee for constructive
comments and suggestions. This work was sponsored by the Faculty
Research (Grant No. FRG2/08-09/070) Hong Kong Baptist University.

\end{document}